# Conserved network motifs allow protein-protein interaction prediction

István Albert[1], Réka Albert [1,2]
1. The Huck Institutes for the Life Sciences, Pennsylvania State University
2. Department of Physics, Pennsylvania State University

ABSTRACT

**Motivation:** High-throughput protein interaction detection methods are strongly affected by false positive and false negative results. Focused experiments are needed to complement the large-scale methods by validating previously detected interactions but it is often difficult to decide which proteins to probe as interaction partners. Developing reliable computational methods assisting this decision process is a pressing need in bioinformatics.
**Results:** We show that we can use the conserved properties of the protein network to identify and validate interaction candidates. We apply a number of machine learning algorithms to the protein connectivity information and achieve a surprisingly good overall performance in predicting interacting proteins. Using a "leave-one-out" approach we find average success rates between 20-50% for predicting the correct interaction partner of a protein. We demonstrate that the success of these methods is based on the presence of conserved interaction motifs within the network.
**Availability:** A reference implementation and a table with candidate interacting partners for each yeast protein are available at http://www.protsuggest.org
**Contact:** iua1@psu.edu

INTRODUCTION

Genome-wide protein interaction maps are among our best models to represent the intricate relations between fundamental biochemical reactions that ultimately define biological processes. High-throughput experimental techniques such as the two-hybrid method have led to the construction of maps for diverse organisms, such as the yeast *S. cerevisiae*, the fruitfly *Drosophila melanogaster* and the nematode worm *C. elegans* (Giot et al. 2003; Li et al. 2004; Uetz et al. 2000). Unfortunately, even comprehensive reconstruction efforts may fail to probe all possible relations, and thus lead to incomplete representations. Moreover, two-hybrid experiments are also strongly affected by false positive results that influence a sizable fraction of the interactions detected this way (von Mering et al. 2002). Focused, small-scale experiments are needed to complement the high-throughput results, but often it is hard to decide what proteins to probe as binding partners for a given protein. In this paper we examine a family of computational approaches that make use of the properties of the known interaction network to predict new interaction candidates. Our proposed methods work by aggregating the conserved network patterns into interaction neighborhoods that proteins belong to. We describe the mechanisms by which this aggregation occurs, evaluate the quality of the predictions for the yeast protein interaction network and provide an implementation available at *http://www.protsuggest.org*

The protein-protein interaction data can be represented as a network whose nodes are proteins, and they are connected by edges if the corresponding proteins interact (Albert and Barabasi 2002). Previous studies have shown that these networks are highly heterogeneous, containing both a large number of proteins with few interaction partners, but also many highly connected "hub" proteins (Jeong et al. 2001; Yook et al. 2004). It has also been shown that certain network motifs such as a triad or tetrad of interactions occur at a significantly higher frequency than that expected from an artificially generated network with similar mathematical properties (Li et al. 2004; Yook et al. 2004). The existence of over-represented sub-networks has been confirmed in a wide variety of complex systems and the conservation of these motifs points to their functional regulatory role (Milo et al. 2004; Milo et al. 2002; Shen-Orr et al. 2002; Wuchty et al. 2003). In this paper we show that we can leverage the information encoded in consensus interaction patterns to generate high relevance predictions for new interaction partners of any given protein. Notably, this performance is achieved without using any kind of prior biological knowledge while at the same time the mechanisms by which the prediction processes take place are readily interpretable.

*Recommender systems*, originally developed for information filtering and E-Commerce applications, are knowledge discovery agents that find preference neighborhoods best fitting past selections and then use that as a seed to search for relevant items (Deshpande and Karypis 2004; Sarwar et al. 2000). A recommender' implementation is not tied to any particular algorithm, it only reflects the operating principles. Great progress has been made in the past few years, and the algorithms and methodologies available today are both accurate and highly scalable. We have chosen a freely available, high-performance recommendation engine developed at the University of Minnesota, named Suggest (http://www-users.cs.umn.edu/~karypis/suggest/index.html) that implements three separate prediction strategies. The methods that we have tested fall into the category of *unsupervised, instance-based learners* that match a new input to patterns found in a training set.

The algorithms implemented in the Suggest engine work by finding instances with similar attributes and then aggregate these attributes to form neighborhoods of similar instances. The prediction candidates are then selected from this neighborhood and ranked with a weighting function. For every prediction new neighborhoods may be formed so that the method may capture individual characteristics. We have chosen to map the protein interactions to this formalism by considering the presence and absence of interactions with every other protein as the attributes of each individual protein. This way the algorithms operate on the connectivity information gleaned from the first and second order neighbors. We expect that interaction motifs that are frequently present within this two-step range will allow us to infer previously hidden information (see Figure 1a). This formalism, coupled with the understanding of how the prediction process takes place, permits us to apply the metaphor of "proteins preferring to interact with certain other proteins". We note here that our concept of an interaction motif captures two important characteristics: 1. a topological layout (triangle, square etc) and 2. the individual node identities within this geometrical entity. For every protein a multitude of network motifs might be present that will generate a large number of candidates, some of which may be present multiple times. In the final step of the prediction the candidates are summed then ranked via a weighing function to create a list of decreasing relevance.

METHODS

**Algorithms**

Each of the three algorithms defines similarity in a different manner, thus providing us with further insight into the properties of the network. The first method is a *signature* driven approach, where a protein's interaction signature (defined as the individual proteins it interacts with) is matched against all known signatures. Each protein is assigned a binary vector of length equal to the number of proteins in the network, *N*, whose nonzero elements indicate the protein signature. Selecting the most similar signatures via a cosine metric, defined as the cosine of their associated vectors' angle in the *N* dimensional space, forms the neighborhood. This method is a variant of the K-nearest neighbors algorithm and is also known as user-based Top-N recommender. The second method, *aggregation*, is based on pre-computing similarities between existing signatures and then selecting interaction candidates from the neighborhoods that contain the proteins that are the interaction partners of the target protein; this method is also known as item-based Top-N recommender. Finally, the third method is a *probabilistic* approach that uses as its similarity measure the conditional probability of an item being present. In particular, the conditional probability that a protein that interacts with **P** also interacts with **Q** is the ratio between the number of proteins that interact with both **P** and **Q** and the number of proteins that interact with **P** (see the Suggest library's documentation for implementation details).

**Interaction patterns**

For each protein in the network, we count the number of interaction patterns that contain it. To effectively compare the density of different motifs, we use the number of patterns per edge pair, defined as the number of interaction patterns divided by $k*(k-1)/2$ where $k$ is the number of neighbors the protein has. This definition is equivalent with the clustering coefficient for triads, and is related to the grid coefficient (Caldarelli et al. 2004) for tetrads. We identify three network motifs that form the basis of correct predictions: triads of proteins with three interactions and protein tetrads with four or five interactions (see Fig. 1b. The number of motifs per edge pair reported on Fig. 2 is the sum of the densities of all three motifs.

RESULTS

We obtained the *S. cerevisiae* protein-protein interaction data from DIP, the Database of Interacting Proteins (Xenarios et al. 2002), containing a total of 4741 proteins and 15409 interactions among them. Since proteins with a single interaction cannot possibly be predicted with the methods that we have set out to evaluate, we iteratively remove these from our network. By the end of the process there were 3394 proteins and 14101 interactions left, with every protein participating in at least two interactions. We test the applicability of three algorithms that for simplicity we denote "*signature*", "*aggregation*" and "*probabilistic*" (for an alternative nomenclature and details see Methods). We use a standard evaluation technique often referred to as "*leave-one-out*" where we place a single interaction in a test set and then use all the remaining data as the training set. We perform such a prediction attempt for every interaction in the database; as each interaction involves two proteins, in total we perform 28202 predictions for each of the three methods. In circumstances such as ours, where the expected number of false positives and false negatives is large, evaluating the overall quality of the predictions poses particular challenges. Many standard procedures such as mean absolute error analysis, confusion matrices or ROC evaluations are ill suited since they are quite sensitive to *a priori* classification errors. Since a rigorous validation process appears to be unfeasible we base our validation methodology upon two observations. 1. The chance of randomly selecting valid partners is at least two orders of magnitude smaller than the probabilities that the algorithms can produce. 2. From a biologist's perspective the existence of an interaction is more important than its non-existence. That is to say, true positives carry more value than true negatives. Therefore we have chosen to evaluate the algorithms solely by their ability to correctly predict existing interactions. The output of the prediction process is a list of candidates, and we will consider the prediction to be correct if the missing interaction is recovered within the candidates. Fortunately there is a way to verify this type of validation by comparing the results obtained for all interactions to those obtained on a high-confidence subset of it. From the 14101 interactions 4871 have either a small scale experimental or paralogous verification (Deane et al. 2002) thus provide us with a "gold standard" that we will use to check to predicted values.

For every method the interaction candidates are ranked by their estimated relevance, thus lower-ranked candidates are expected to be less accurate. The most stringent test of the algorithm is the fraction of cases when the first candidate represents a successful prediction. We find that this happens in approximately 8% of the cases. In other words, in 1127 instances we obtain the correct prediction as our first guess! As we increase the size of the candidate list, the rate of success increases, at the expense of generating false positives. On Figure 2 we compare the performance of the three recommender methods as a function of the size of the candidate set. Note that the rate of increase is highly nonlinear, tapering off at a higher number of candidates. This indicates that the methods are accurate enough to produce their best candidates in the first few returns. The frequency of the correct predictions allowing just the first five candidates is around 20% for every method.

An interaction is defined between two interaction partners, and whenever these partners correspond to different proteins the interaction can be predicted from both "ends", that is using the patterns representing either protein. If we consider a prediction to be successful if any one of the two proteins correctly predicts it then the overall prediction quality becomes notably higher. This prediction quality is not directly comparable to one-directional predictions since they may correspond to a larger number of candidates for each prediction. The resulting values are, however, significantly better even when compared to the values corresponding to twice as many candidates in the unidirectional predictions. This again points to the ability to return relevant values within the first candidates. On the other extreme, a prediction could be defined to be valid only in full consensus when both "ends" agree. This decreases the sensitivity of the methods, as they can only find 10% of the interactions but increases their precision, generating a correct answer in more than 50% of the cases.

The quality measure in Figure 2 refers to the ability to predict a missing interaction with respect to all interactions in the network. Since the number of interactions per protein follows a scale-free distribution with a high variance, there is another quality measure of interest, the ability to predict missing interactions for a certain protein. As we will see later, this measure strongly correlates with the number of network motifs that a protein participates in. Overall for a candidate pool size of 5 we were able to generate at least one correct prediction for 40% of the proteins. The most influential factor in whether a protein can be predicted for at all appears to be its node degree (the number of interactions a protein participates in) with more than 85% of the "unpredictable" proteins having less than 5 interactions (see Figure 3a). This is well within reason, as the fewer the local connections, the less information the algorithm is able to use for further inference. For the proteins with at least one successful prediction, we define the protein prediction quality as the percentage of correctly predicted interactions for the given protein. We find that the average prediction quality among the approximately 1500 "predictable" proteins is a high 42%. As shown on Figure 3b, prediction qualities between 20-50% are approximately equiprobable, with less frequency for the low and high-end values. The successful predictions accounted for more than two-thirds of the total number of interactions. When we average the predictability of proteins with given degree we see that this value holds approximately for the majority of protein degrees. At the two extremes, however, at the very low (less than 4) and at the very high (more than 40) degrees we observe different behaviors. The quality of prediction for low values tends to be higher while the prediction quality for the high connectivity nodes tends to be lower than this average.

In conclusion, three conceptually different methods lead to remarkably consistent predictions that, without taking into account any of the biological characteristics of the proteins and their interactions, are surprisingly successful in predicting missing interactions. The key to this success is necessarily in the topology of the network of interactions, and we find it is rooted in the local interaction patterns between proteins. To illustrate the idea, let us focus on the probabilistic method. The algorithm judges the relevance of an interaction between two nodes by determining the ratio between the number of nodes that connect to both and the number of nodes that connect to one of them. This means that if three nodes are connected in a triangle, the probabilistic method may be able to predict any of the edges based on the existence of the other two. The existence of the triangle is therefore a necessary but not sufficient condition for a successful prediction. Thus the number of triangles in the first neighborhood of a protein (also known as a node's clustering coefficient, (Watts and Strogatz 1998) ) is expected to correlate with the predictability of its edges. We find that the smallest interaction patterns that convey high predictability for all methods are square motifs with four nodes and four or five edges (see Fig. 1b).

We determined the density of these three motifs in the neighborhood of each protein, and correlated it with the frequency of successful predictions of its interactions (see Methods). Figure 3 presents a scatterplot of motif density/protein predictability pairs for each protein, as well as the average motif density of proteins corresponding to a certain predictability. The figure clearly indicates that the quality of predictions increases with motif density. An average of one network motif per edge pair leads to an impressive 40% success rate, thus a high frequency of interaction patterns ensures high edge predictability. We have verified that the converse is also true, and the absence of interaction motifs leads to unpredictability.

Using this new knowledge on their role, the presence of interaction motifs can be explicitly leveraged in the prediction process to lead to a much higher success rate. On Fig. 2b we illustrate the performance of an optimized algorithm that uses known information about the network. Since the existence of an interaction can be predicted from two proteins, we choose to use the candidates generated from the neighborhood of the protein with the higher motif density. We can also ensure that during evaluation we are not trying to reproduce false-positives by generating predictions only for the pool of high-confidence interactions cross-validated by one or more methods (Deane et al. 2002). We obtain encouraging results showing that the success rate of the high-confidence predictions increases to an impressive 50% on a 5 candidate variant.

DISCUSSION

Several methods of mining the proteome data have been proposed in the literature. Supervised Bayesian Learners have been used to combine multiple genomic features into reliable predictions of interacting proteins (Jansen et al. 2003). Inference rules for new protein functions were formed by combining known protein features with interaction partner information (Oyama et al. 2002). A Markov Random Field formalism was successfully applied to predicting protein function based on the local functional density of interacting neighbors (Letovsky and Kasif 2003) and protein-protein interaction sites could be identified by using the profiles of spatially or sequentially neighboring sequences (Koike and Takagi 2004). What most separates our approach from the previously published results is that the recommendation algorithms that we employ work without requiring additional biological knowledge and use the connectivity data as the only source of implicit information.

In the present paper we tested the applicability of a group of recommender systems to predicting protein-protein interactions. The methods that we describe are freely available and work remarkably well, and our results indicate that they have the potential to be a valuable addition to other bioinformatics methods. We were able to correctly predict a high percentage of the interactions in a protein interaction network. We have shown that the success of the algorithm is rooted in the abundance of conserved interaction patterns in the network. As such conserved motifs have been reported in various other biological networks, we anticipate a wide applicability of recommender methods to predict unknown interactions between cellular components. The three methods that we investigated exhibit different performances, yet it would be premature to conclude that either of them is necessarily better than the rest. Previous work (McNee et al. 2002) comparing prediction strategies in the realm of information retrieval has demonstrated that quality measures do not properly capture the applicability of the predictions with respect to user tasks and goals. They found that some methods were biased towards finding "ground truths" while others were more sensitive to local similarities and led to serendipitous discoveries. In our case we observed a significant overlap between the candidates generated by different methods yet the rank of the candidates varied substantially. We believe that further studies are needed to investigate the nature of predictions and the mechanisms that govern them.

To demonstrate the utilization of these algorithms we implement them as an interactive web tool available at http://www.protsuggest.org. The web-site operates on the yeast protein interaction network obtained from DIP, and generates interaction candidates based on a list of interaction partners entered in a query window. We also offer for download a file containing the first 25 most likely interaction candidates of each protein. We expect that incorporating information about protein structure or functional classification in the prediction phase may significantly enhance the quality of the predictions. For example the wrapping of the hydrogen backbone bonds has been show to correlate with the interactivity of individual domains (Fernandez et al. 2004) while in a different study, functional groups have been found in hidden topological structures (Bu et al. 2003). The effects of integrating several types of data will be explored in future work. Recommender systems are widely deployed in e-commerce and information filtering systems, from Amazon's book matching engine to Yahoo Launch's music recommendation services. Our results demonstrate that biologists should also start harvesting their power.

ACKNOWLEDGEMENTS

We thank George Karypis for help and advice regarding the usage of the SUGGEST library.

FIGURE LEGENDS

**Figure 1** a). Illustration of a network-motif-based prediction process. This hypothetical sub-network can be decomposed into six squares starting with XAYB etc. Within these squares only the presence of A, B and C are conserved. This network motif will generate C as an interaction candidate for any protein that forms a square with A, B, and any of X,

Y or Z. b) The smallest network motifs that can recover missing information. A simple triangle pattern of first order neighbors gives the correct prediction only with the probabilistic method, but is successful with any of the three if one of the nodes contains a self-interaction. In a square with four nodes (first and second order neighbors) and four edges any edge can be predicted based on the three others. Every edge is highly predictable in a double triangle of four nodes and five edges.

**Figure 2** Interaction prediction success rate as a function of the number of candidates generated. a) Success rate of predictions from one endpoint of the interaction in the test set. b) We can obtain an improved success rate if the prediction algorithm and comparison set are optimized. In this example we use the candidates predicted by the endpoint with higher motif density, and compare only with high-confidence interactions.

**Figure 3.** Quantifying proteins by the predictability of their interactions. a) Histogram of the degree of proteins whose interactions could not be predicted by a five-candidate algorithm. More than 85% of these "unpredictable" proteins have less than 5 interactions. b) Histogram of successfully predicted proteins with respect to their prediction quality (percentage of their correctly predicted interactions). Quality levels between 20-50% are approximately equally distributed across methods, while low and high qualities are more infrequent.

**Figure 4.** Correlation between protein predictability and local network motif density. a) Scatterplot of number of motifs per edge pair (see Methods) versus protein prediction quality for each protein. The prediction quality denotes the frequency of correct 5-candidate predictions for each protein, the distribution is shown for the probabilistic method. b) The same data averaged over prediction quality bins of size 10%, this time for all three methods. We choose to average over prediction quality to ensure good statistics in each bin.

FIGURES

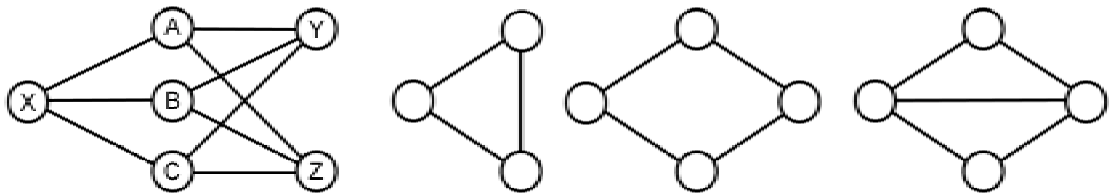

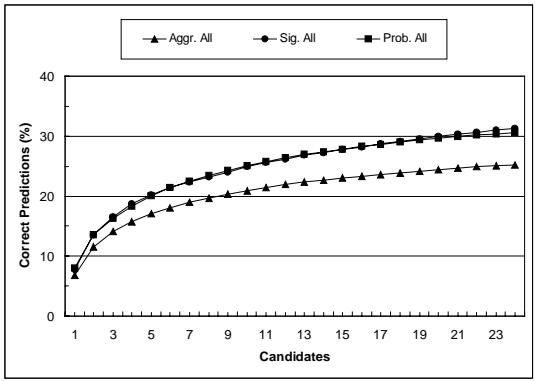 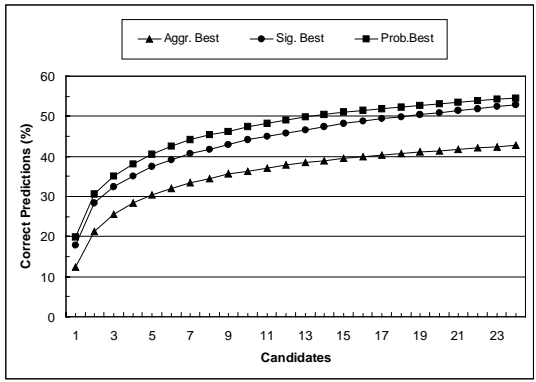

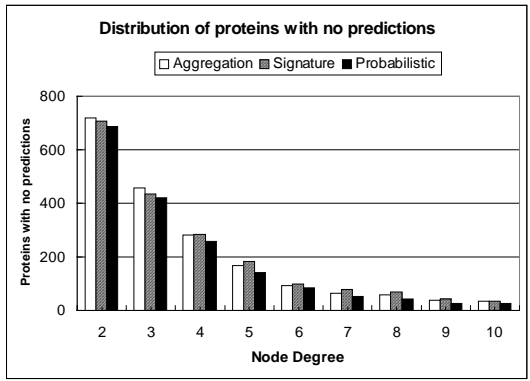 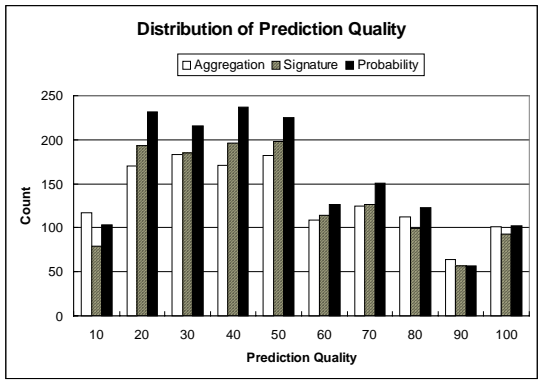

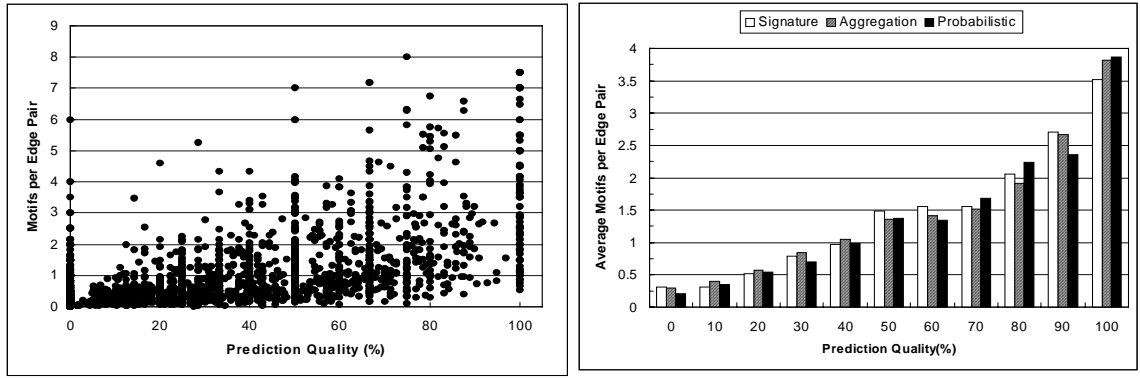

# WEBSITE REFERENCES

1. http://dip.doe-mbi.ucla.edu/, the Database of Interacting Proteins (DIP).
2. http://www-users.cs.umn.edu/~karypis/suggest/index.html, the SUGGEST recommendation library.
3. http://www.protsuggest.org, web interface to the methods presented in the paper.